\documentclass[prl,aps,twocolumn,groupedaddress,floats,showpacs,final]{revtex4-1}

\usepackage{graphicx}
\usepackage{bm}
\usepackage{color}
\begin{document}

\title{Universal Properties of the Higgs Resonance in (2+1)-Dimensional $U(1)$ Critical Systems}
\author {Kun Chen$^{1,2}$, Longxiang Liu$^{1}$}
\author {Youjin Deng$^{1,2}$}
\email{yjdeng@ustc.edu.cn}
\author {Lode Pollet$^{3}$}
\email{lode.pollet@physik.uni-muenchen.de}
\author {Nikolay Prokof'ev$^{2,4}$ }
\email{prokofev@physics.umass.edu}
\affiliation{$^{1}$ National Laboratory for Physical Sciences at Microscale and Department of Modern Physics, University of Science and Technology of China, Hefei, Anhui 230026, China}
\affiliation{$^{2}$Department of Physics, University of Massachusetts, Amherst, Massahusetts 01003, USA}
\affiliation{$^{3}$Department of Physics and Arnold Sommerfeld Center for Theoretical Physics, Ludwig-Maximilians-Universit{\"a}t M{\"u}nchen, D-80333 M{\"u}nchen, Germany}
\affiliation{$^{4}$Russian Research Center ``Kurchatov Institute", 123182 Moscow, Russia }
\date{\today}

\begin{abstract}
We present spectral functions for the magnitude squared of the order parameter
in the scaling limit of the two-dimensional superfluid to Mott insulator quantum phase transition at constant density, 
which has emergent particle-hole symmetry and Lorentz invariance. The universal functions for the superfluid, 
Mott insulator, and normal liquid phases reveal a low-frequency resonance which is relatively sharp and is followed by a 
damped oscillation (in the first two phases only) before saturating to the quantum critical plateau. 
The counterintuitive resonance feature in the insulating and normal phases calls for deeper understanding 
of collective modes in the strongly coupled (2+1)-dimensional relativistic field theory.  
Our results are derived from analytically continued correlation functions obtained from path-integral Monte 
Carlo simulations of the Bose-Hubbard model.
\end{abstract}


\maketitle

Field theories of a complex scalar order parameter, $\Psi$, can have two types of collective excitations.
The first one originates from fluctuations of the  phase of $\Psi$ and describes a Bogoliubov sound mode.
The second one, if present, describes amplitude fluctuations and is associated with a Higgs mode.
In superfluids, sound excitations are gapless while the Higgs mode, if present, is gapped
but the gap may go to zero under special circumstances such as an emergent particle-hole
symmetry and Lorentz invariance.
This is what happens in the vicinity of the superfluid (SF) to Mott insulator (MI)
quantum critical point (QCP) of the Bose-Hubbard model when the phase transition is crossed at
constant density.

Mean-field theory predicts a stable Higgs particle. In (3+1) dimensions, where the QCP
is a Gaussian fixed point (with logarithmic UV corrections), there is compelling experimental
evidence for the existence of a Higgs mode, most beautifully illustrated for the TlCuCl$_3$ compound~\cite{ruegg} (see Ref.~\cite{Bissbort} for the latest results with cold gases).
In (2+1) dimensions, where scaling theory is expected to apply, the massive Higgs particle is
strongly coupled to sound modes and it was argued for a long time, on the basis of a
$1/N$ expansion to leading order ($N=2$ corresponds to our case), that it cannot survive
near criticality~\cite{chubukov,sachdev,zwerger,book}.
Moreover, since the longitudinal susceptibility diagram has an IR
divergence going as $\omega^{-1}$, it may well dominate any possible Higgs peak.
However, it was recently emphasized that the type of the probe is important~\cite{pod11, huber, huber2}:
for scalar susceptibility (i.e., the correlation function of $\vert \psi \vert^2$) the spectral function
$S(\omega)$ vanishes as $\omega^3$ at low frequencies~\cite{chubukov,pod11}, and this offers better conditions 
for revealing the Higgs peak. In the scaling limit the theory predicts that $S(\omega )$ in the SF phase takes the form
\begin{equation}\label{scalinginsf}
S_{\rm SF}(\omega) \propto \Delta^{3-2/\nu}\Phi_{ \rm SF} (\frac{\omega}{\Delta}) \;,
\end{equation}
where $\Delta$ is the MI gap for the same amount of detuning from the QCP,
and $\nu=0.6717$ is the correlation length exponent for the $U(1) \equiv O(2)$ universality in $(2+1)$ dimensions~\cite{Machta,Vicari}.
The universal function $\Phi_{\rm SF} (x)$ starts as $\Phi_{\rm SF} (x\to 0) \propto x^3$ and saturates to a quasiplateau
$\Phi_{\rm SF} (x \gg 1 ) \propto x^{3-2/\nu} \approx x^{0.0225}$.
The Higgs resonance (at $x\sim 1$) can be seen right before the incoherent
quantum critical continuum with weak $\omega$ dependence.

We are not aware of solid state studies of the Higgs mode in two-dimensional (2D) superfluids near the QCP.
Recently, the cold atom experiment~\cite{bloch}, where a 2D Bose-Hubbard system was gently "shaken" by modulating
the lattice laser intensity and probed by \emph{in situ} single site density measurements, saw a broad spectral response
whose onset softened on approach to the QCP, in line with the scaling law (\ref{scalinginsf}), and no Higgs resonance.
%The experiment was explained by a heuristic combination of Gutzwiller mean-field theory and
%the arguments presented in Ref.~\cite{pod11}, which contradict scaling theory (note that the
%artificially predicted sharpening of the peak for larger detuning was not seen experimentally).
This outcome can be explained by tight confinement, finite temperature, and detuning from the QCP,
as shown by quantum Monte Carlo (MC) simulations~\cite{pollet} performed for the
experimental setup "as is" in the spirit of the quantum simulation paradigm~\cite{lode_review}.
On the other hand, simulations for the homogeneous Bose-Hubbard model (below, $J$, $U$, and $\mu$
stand for the tunneling amplitude, on-site interaction, and chemical potential, respectively;
in what follows energy and frequency are measured in units of $J$)
\begin{equation}\label{eq:BH}
H = - J \sum_{<ij>}  b_i^\dag b_j^{\,}
+ \frac{U}{2} \sum_i n_i(n_i-1) - \mu \sum_i n_i  \, ,
\end{equation}
in the vicinity of the SF-MI point featuring emergent particle-hole symmetry and Lorentz invariance \cite{fisher89}
unambiguously revealed a well-defined Higgs resonance which becomes more pronounced on approach to the QCP \cite{pollet}.
However, its universal properties, {\it i.e.}, the precise structure of $\Phi_{\rm SF}(x)$, were not answered in Ref.~\cite{pollet}.

The Higgs mode is not discussed in the MI phase since the order parameter is zero in the thermodynamic limit.
Likewise, no resonance is expected in the normal quantum critical liquid (NL), {\it i.e.},
at finite temperature for critical parameters $(U,\mu)=(U_c,\mu_c)$. However, simulations 
reveal a resonance in the MI phase right after the gap threshold~\cite{pollet} suggesting that 
finite-energy probes are primarily sensitive to local correlations at length scales where MI and 
SF are indistinguishable. The universality of the MI response was likewise never clarified.

In their most recent calculation, Podolsky and Sachdev \cite{Podev} found that including
next-order corrections in a $1/N$ expansion in the scaling limit radically changes previous
conclusions in that $S(\omega)$ does contain an oscillatory component, in line with
MC simulations. However, the precise shape of the $\Phi_{\rm SF}(x)$ function could not
be established within the approximations used.

\begin{figure}[htbp]
\includegraphics[scale=0.4,angle=0,width=0.9\columnwidth]{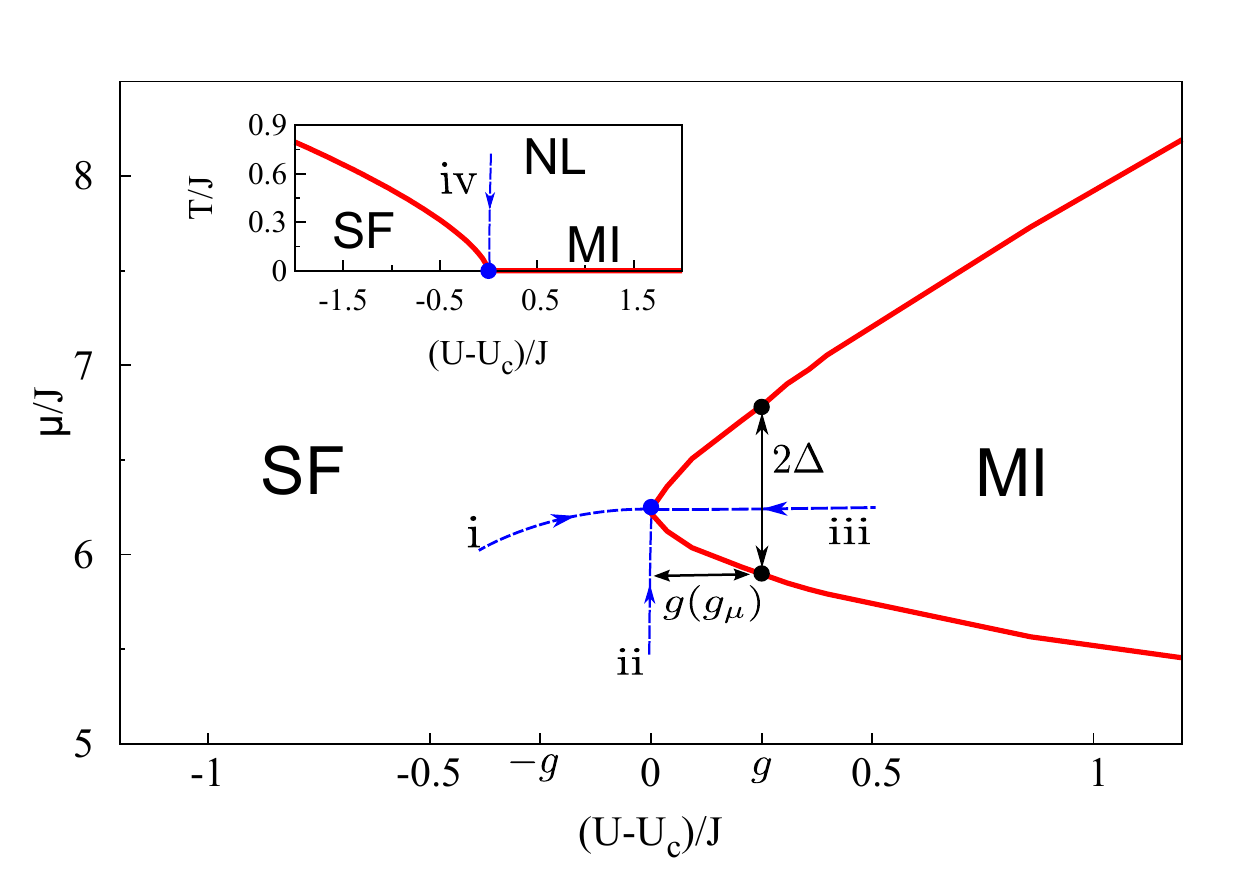}
\caption{\label{fig:phasediag} (Color online) Ground state phase diagram of the  Bose-Hubbard model in the vicinity of the QCP marked by a large (blue) dot (based on Ref.~\cite{soyler} data).
% The full (red) lines denote the MI boundaries from Ref.~\cite{soyler}.
The (blue) dashed curves specify trajectories in parameter space used to detune the system away from the QCP (trajectories {\romannumeral 1} and {\romannumeral 3} correspond to unity filling factor $n=1$, trajectory {\romannumeral 2} has constant interaction strength).
The (black) lines with arrows explain how the characteristic energy scale $\Delta $
is obtained for these parameters (see text).
The inset shows the phase diagram at finite temperature, and the trajectory taken in the NL phase. }
\end{figure}

\begin{figure}[htbp]
\includegraphics[scale=0.4,angle=0,width=0.9\columnwidth]{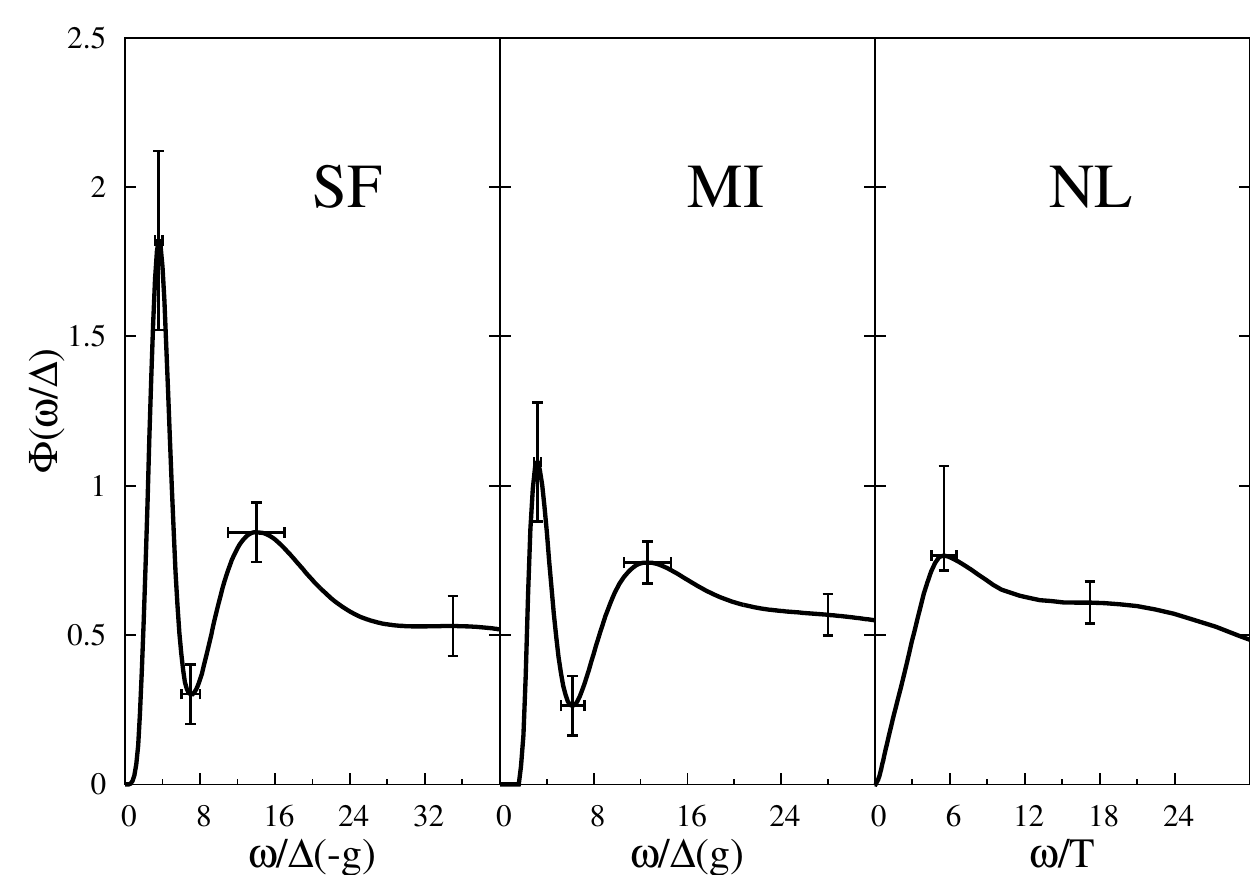}
\caption{\label{fig:main} Universal spectral functions for scalar response in the superfluid,
Mott insulator, and normal liquid phases. For SF, the Higgs peak is at $\omega_H/\Delta=3.3(8)$, for MI, $\omega_H/\Delta=3.2(8)$, and for NL, $\omega_H/T=6(1)$. There is a secondary peak around
$\omega/\Delta \approx 15$ in the SF and MI phases, and all responses reach a quasiplateau at the
same height $0.6(1)$ at higher frequencies.
The error bars on $\Phi_{\rm SF, \rm MI} $ come from the spread of collapsed curves, while the ones on $\Phi_{\rm NL}$ 
are based on the variance of the analytical continuation results \cite{pollet}.}
%\textcolor{red}{ $\Phi_{\rm SF, \rm MI} $'s error bars are estimated by the collapse of spectral function with different parameters, while $\Phi_{\rm NL}$'s are obtained from different analytical continuation approaches. }}
\end{figure}

In this Letter, we aim to determine the universal scaling spectral functions when approaching the
QCP from the SF, MI, and NL phases.
%(also known as renormalized classical or quantum disordered) phases, respectively.
We rely on the worm algorithm~\cite{worm,worm2, worm_lode} in the path integral representation
to perform the required large-scale simulations.
By collapsing spectral functions evaluated along the trajectories specified by the dashed lines in Fig \ref{fig:phasediag},
we extract universal features for all three phases.
They are  summarized in Fig.~\ref{fig:main}, which is our main result.
Surprisingly, all of them include a universal resonance peak (relatively sharp in SF and MI phases),
followed by a broad secondary peak (in SF and MI phases only) before merging with the incoherent critical quasiplateau
(the plateau value is the same in all cases, as expected).
Our results are in agreement with scaling theory, and firmly establish that the damped resonance
is present in all three phases. (The integrated spectral weight of the $\omega^3$ law at low frequencies 
is too small to be resolved reliably by analytical continuation methods \cite{pollet}.)
%%%%%%%%%%%%%%%%%%%%%%%%%%%%%%%%%%%%%%%%%%%%%%%%%%%%%%%%%%%%%
%%%%%%%%%%%%%%%%%%%%%%%%%%%%%%%%%%%%%%%%%%%%%%%%%%%%%%%%%%%%%%

The phase diagram of the 2D Bose-Hubbard model, shown in Fig. \ref{fig:phasediag},
is known with high accuracy \cite{monien,soyler,soyler2} at both zero and finite temperature.
The QCP is located at $U_c=16.7424(1)$, $\mu_c=6.21(2)$.
When the system is slightly detuned from the QCP, either by changing the chemical potential
or the interaction strength, we define the corresponding characteristic energy scale $\Delta$
using the energy gap in the MI phase, $E_{\rm gap}(g)$, by the rule illustrated
in Fig.~\ref{fig:phasediag}: For positive $g=(U-U_c)/J$ it is half the gap,
$\Delta (g>0) = E_{\rm gap}(g)/2$, where $E_{\rm gap}=\mu_c^{(+)} - \mu_c^{(-)}$
is deduced from the upper and lower critical chemical potentials for a given $g$.
For $g<0$ along the trajectory {\romannumeral 1} in the SF phase it is $\Delta(g<0)=E_{\rm gap}(-g)/2$.
For $U=U_c$ and negative $g_{\mu} = (\mu-\mu_c)/J$ along the trajectory {\romannumeral 2} in the SF phase
we first find $g$ such that $\mu_c^{(-)}(g)=\mu$ and then define $\Delta (g_{\mu} ) = C E_{\rm gap}(g)/2$
where the constant $C=1.2$ (see below) is fixed by demanding that the universal function is the same along both SF trajectories. Note that $E_{\rm gap}(g)$ in the thermodynamic limit can be determined accurately
from the imaginary time Green function data~\cite{soyler} and finite-size scaling analysis.
%%%%%%%%%%%%%%%%%%%%%%%%%%%%%%%%%%%%%%%%%%%%%%%%%%%%%%%%%%%%%
%%%%%%%%%%%%%%%%%%%%%%%%%%%%%%%%%%%%%%%%%%%%%%%%%%%%%%%%%%%%%%

To study the scalar response, we can imagine adding a small uniform modulation term to the Hamiltonian
% Eq.(\ref{eq:BH}),
\begin{equation}\label{modulation}
\delta H(t) =- \delta J \cos (\omega t) \sum_{<ij>}  b_i^\dag b_j^{\,}  \equiv \frac{\delta J}{J} \;K(t) \;,
\end{equation}
where $\delta J/J \ll 1$. The imaginary time correlation function for kinetic energy,
$\chi(\tau)=\langle K(\tau)K(0) \rangle-\langle K \rangle^2$, is related to $S(\omega)$ through the
spectral integral with the finite-temperature kernel,
%$N(\tau,\omega)=\exp(-\omega \tau)+\exp(-\omega(\beta-\tau))$,
$N(\tau,\omega)=e^{-\omega \tau}+e^{-\omega(1/T-\tau)}$:
\begin{equation}\label{ancon}
\chi(\tau)=\int_{0}^{+\infty} N(\tau,\omega) S(\omega) \;.
\end{equation}
We employ the same protocol of collecting and analyzing data as in Ref.~\cite{pollet}. More specifically,
in the MC simulation we collect statistics for the correlation function at Matsubara frequencies
$\omega_n=2\pi Tn$ with integer $n$
\begin{equation}\label{chi}
\chi(i\omega_n)= \langle K(\tau)K(0)\rangle_{i\omega_n}+\langle K \rangle
\end{equation}
which is related to $\chi(\tau)$ by a Fourier transform.
In the path integral representation, $\chi(i\omega_n)$ has a direct unbiased estimator, $\vert \sum_k e^{i\omega_n \tau_k} \vert^2$,
where the sum runs over all hopping transitions in a given configuration, i.e. there is no need to add 
term (\ref{modulation}) to the Hamiltonian explicitly.
Once $\chi(\tau)$ is recovered from $\chi(i\omega_n)$, the analytical continuation methods 
described in Ref.~\cite{pollet} are applied to extract the spectral function $S(\omega)$.
A discussion on the reproducibility of the analytically continued results for this type of 
problem can also be found in Ref.~\cite{pollet}.

We consider system sizes significantly larger than the correlation length by a factor
of at least 4 to ensure that our results are effectively in the thermodynamic limit.
Furthermore, for the SF and MI phases, we set the temperature $T=1/\beta$ to be much smaller
than the characteristic Higgs energy, so that no details in the relevant energy part of the spectral
function are missed.

\begin{figure}[htbp]
\includegraphics[scale=0.4,angle=0,width=0.9\columnwidth]{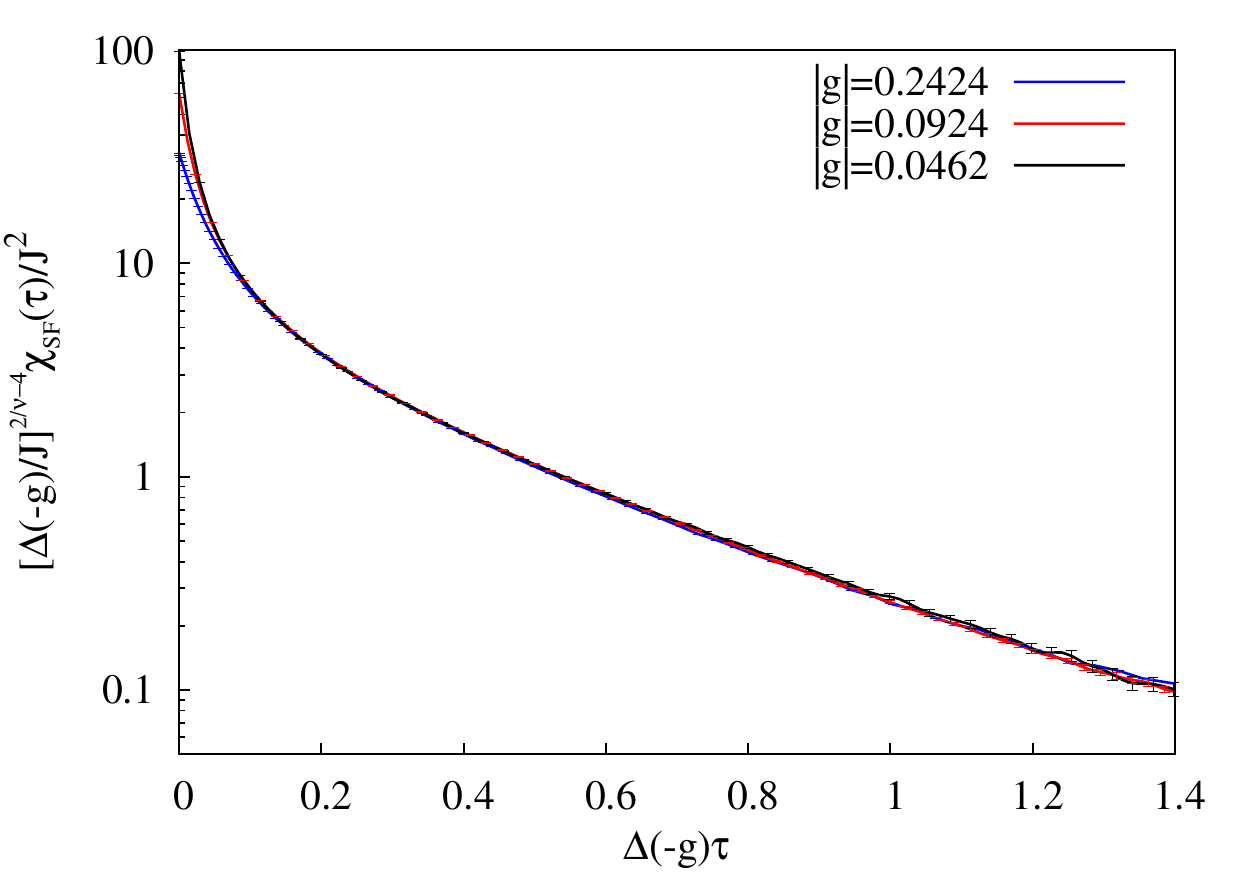}
\caption{\label{fig:tau}(Color online) Collapse of correlation functions in imaginary time domain 
for different values of $U$ along trajectory {\romannumeral 1} in the SF phase, labeled by the detuning $g=(U-U_c)/J$.
}
\end{figure}

\begin{figure}[htbp]
\includegraphics[scale=0.4,angle=0,width=0.9\columnwidth]{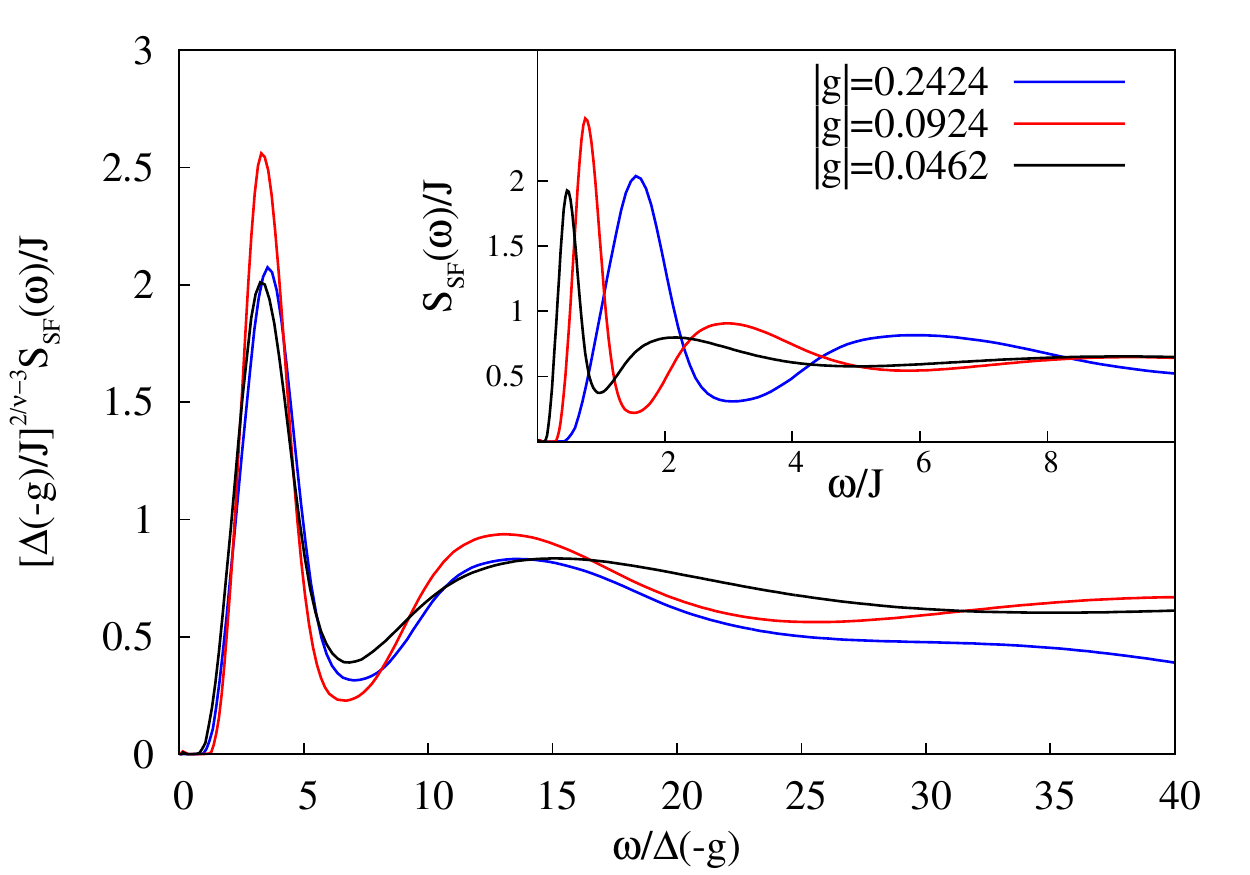}
\caption{\label{fig:diff_u}(Color online) Collapse of spectral functions for different values of $U$
along trajectory {\romannumeral 1} in the SF phase, labeled by the detuning $g=(U-U_c)/J$.
Inset: Original data for $S_{\rm SF}(\omega )$.}
\end{figure}

We consider two paths in the SF phase to approach the QCP:
by increasing the interaction $U \rightarrow U_c$ at unity filling factor $n=1$
(trajectory {\romannumeral 1} perpendicular to the phase boundary in Fig \ref{fig:phasediag}),
and by increasing $\mu \rightarrow \mu_c$ while keeping $U=U_c$ constant
(trajectory {\romannumeral 2} tangential to the phase boundary in Fig \ref{fig:phasediag}).
We start with trajectory {\romannumeral 1} by considering three parameter sets for $(|g|,L,\beta )$:
$(0.2424, 20, 10)$, $(0.0924, 40, 20)$, and $(0.0462, 80, 40)$.
The prime data in imaginary time domain are shown in Fig.~\ref{fig:tau} using scaled
variables to demonstrate collapse of  $\chi(\tau)$ curves at large times. 
Analytically continued results are shown in the inset of Fig.~\ref{fig:diff_u}.
After rescaling results according to Eq.~(\ref{scalinginsf}), we observe data collapse
shown in the main panel of Fig.\ref{fig:diff_u}.
This defines the universal spectral function in the superfluid phase $\Phi_{\rm SF}$.
%%%%%%%%%%%%%%%%%%%%%%%%%%%%%%%%%%%%%%%%%%%%%%%%%%%%%%%%%%%%%
%%%%%%%%%%%%%%%%%%%%%%%%%%%%%%%%%%%%%%%%%%%%%%%%%%%%%%%%%%%%%%
%%%%%%%%%%%%%%%%%%%%%%%%%%%%%%%%%%%%%%%%%%%%%%%%%%%%%%%%%%%%%
%%%%%%%%%%%%%%%%%%%%%%%%%%%%%%%%%%%%%%%%%%%%%%%%%%%%%%%%%%%%%%

\begin{figure}[htbp]
\includegraphics[scale=0.4,angle=0,width=0.9\columnwidth]{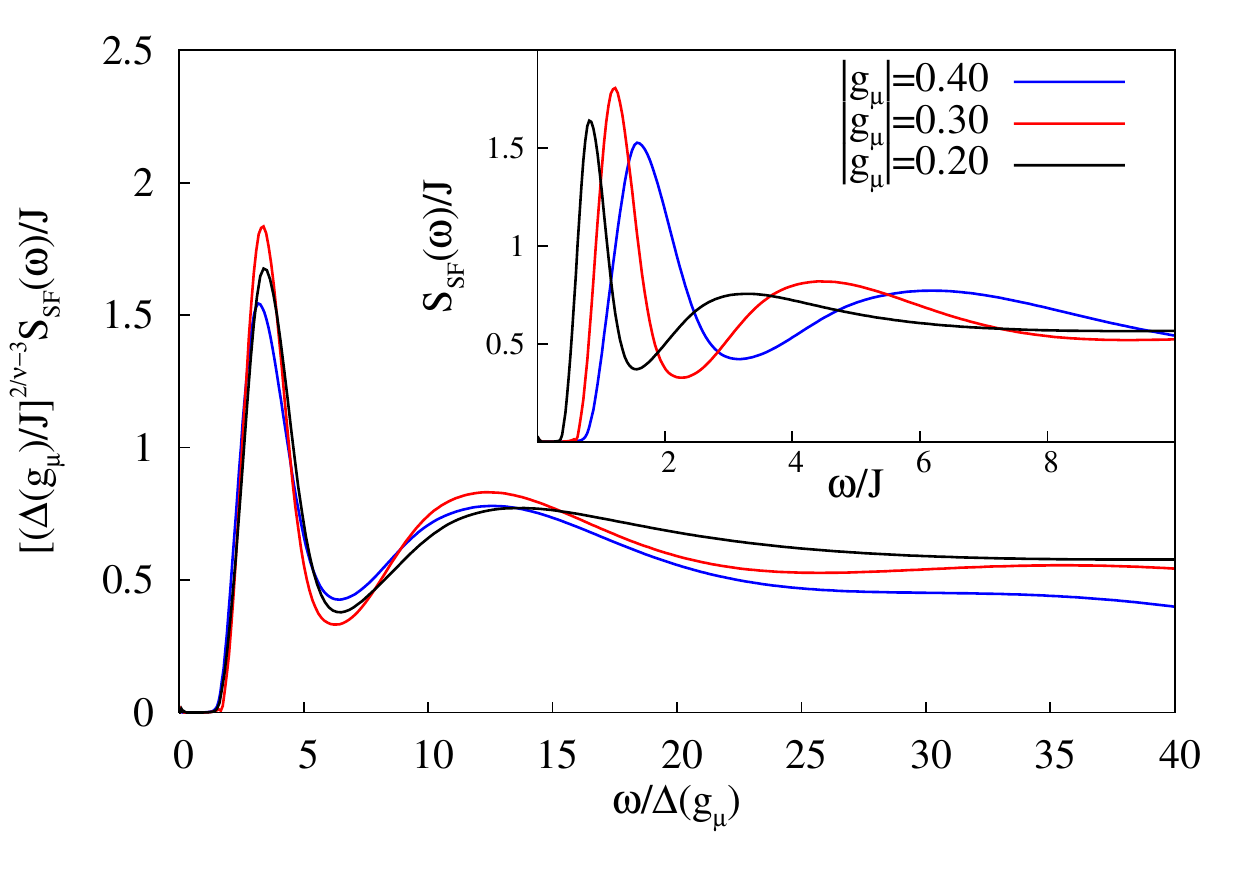}
\caption{\label{fig:diff_mu}(Color online) Collapse of spectral functions for different
$\mu$ along trajectory {\romannumeral 2} in the SF phase, labeled by the detuning
$g_{\mu}=(\mu-\mu_c)/J$. Inset: original data for $S_{\rm SF}(\omega )$.}
\end{figure}

When approaching the QCP along trajectory {\romannumeral 2}, with $(|g_{\mu}|,L,\beta ) =
(0.40, 25, 15)$,  $(0.30, 30, 15)$,  and $(0.20, 40, 20)$ we observe a similar data collapse
and arrive at the same universal function $\Phi_{\rm SF}$; see Fig. \ref{fig:diff_mu}.
The final match is possible only when the characteristic energy scale
$\Delta (g_{\mu})= C \Delta (g(g_{\mu}))$ involves a factor of $C=1.2$.

%%%%%%%%%%%%%%%%%%%%%%%%%%%%%%%%%%%%%%%%%%%%%%%%%%%%%%%%%%%%%%
%%%%%%%%%%%%%%%%%%%%%%%%%%%%%%%%%%%%%%%%%%%%%%%%%%%%%%%%%%%%%%

The universal spectral function $\Phi_{\rm SF}$ has three distinct features: a) A pronounced peak at $\omega_H/\Delta \approx 3.3$,
which is associated with the Higgs resonance. Since the peak's width $\gamma /\Delta \approx 1$ is comparable to its energy, the
Higgs mode is strongly damped. It behaves as a well-defined particle only in a moving reference frame;
b) A minimum and another broad maximum between $\omega /\Delta \in [5,25]$ which may originate from
multi-Higgs excitations \cite{pollet}; c) the onset of the quantum critical quasiplateau,
in agreement with the scaling hypothesis (\ref{scalinginsf}), starting at $\omega/\Delta \approx 25$.
These features are captured by an approximate analytic expression with normalized $\chi^2 \sim 1$,
\begin{equation}\label{universalfunc}
\Phi_{\rm SF}(x)=\frac{0.65x^3}{35+x^{2/\nu}}\left[ 1+\frac{7 \sin (0.55x)}{1+0.02x^3}\right]
\end{equation}
We only claim that a plateau is consistent with our imaginary time data
and emerges from the analytic continuation procedure which seeks smooth spectral functions; 
{\it i.e.,} other analytic continuation methods may produce an oscillating behavior 
in the same frequency range within the error bar in Fig.~\ref{fig:main} 

\begin{figure}[htbp]
\includegraphics[scale=0.4,angle=0,width=0.9\columnwidth]{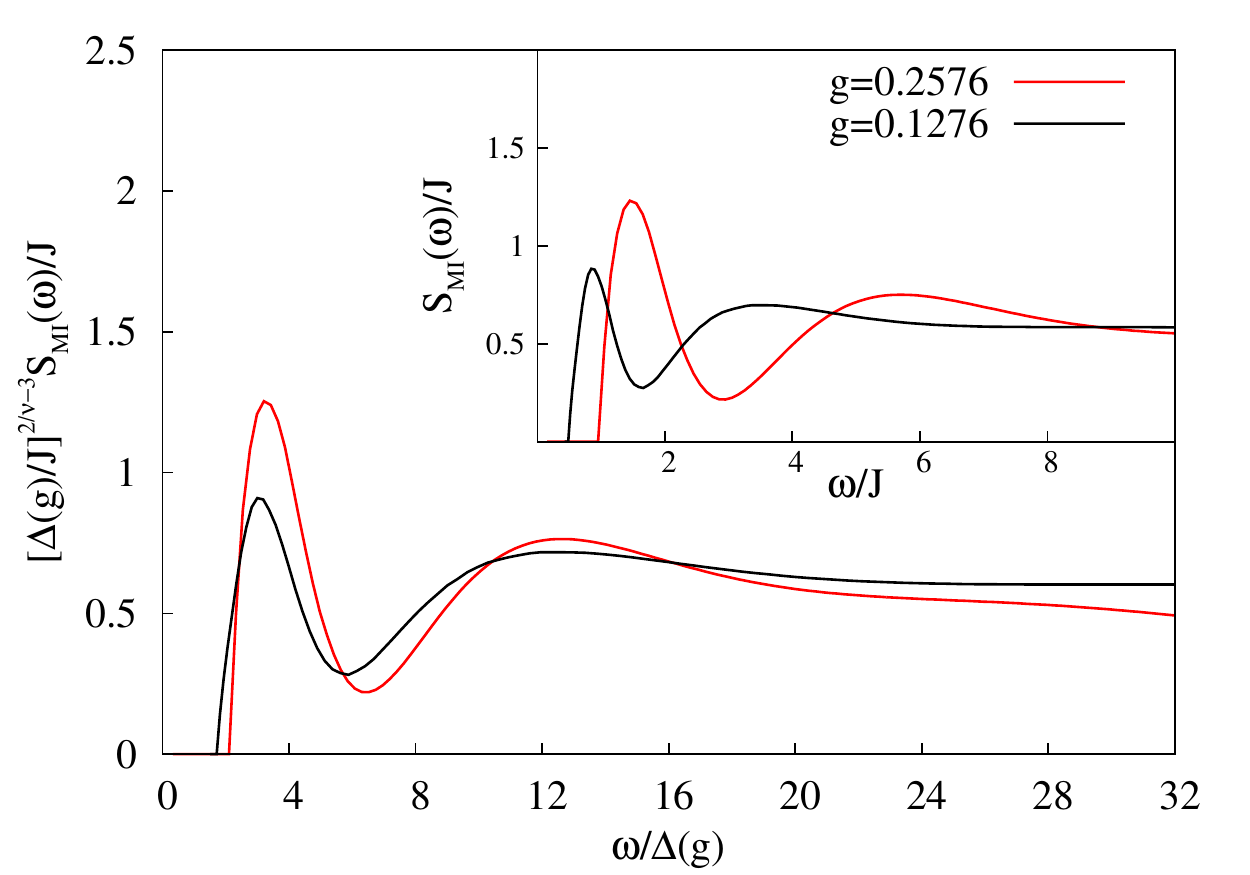}
\caption{\label{fig:diff_u_mi}(Color online) Collapse of the spectral functions
for different $U$ along trajectory {\romannumeral 3} in the MI phase,
labeled by the detuning $g=(U-U_c)/J$. Inset: Original data for $S_{\rm MI}(\omega )$.}
\end{figure}

In the MI phase we approach the QCP along trajectory {\romannumeral 3} in Fig \ref{fig:phasediag}.
The scaling hypothesis for the spectral function has a similar structure to the one in Eq.~(\ref{scalinginsf}),
\begin{equation}\label{scalinginmi}
S_{\rm MI}(\omega) \propto \Delta^{3-2/\nu}\Phi_{\rm MI}(\frac{\omega}{\Delta}) \;.
\end{equation}
The low-energy behavior of $\Phi_{\rm MI}$ starts with the threshold singularity at the particle-hole gap value,
$\Phi_{\rm MI}(x) \approx 1/\log^2(4/(x-2))\theta(x-2)$, see Ref.~\cite{Podev}. At high frequencies
$\Phi_{\rm MI}(x\gg 1)$ has to approach the universal quantum critical quasiplateau
(same as in the SF phase). Our results for the spectral functions at $g=0.2576$ (with $L=20,\beta=10$) and $g=0.1276$  (with $L=40,\beta=20$) are presented in Fig.\ref{fig:diff_u_mi}.
The universal scaling spectral function shows an energy gap
(this is also fully pronounced in the imaginary time data).
The left-hand side of the first peak is much steeper than in the SF phase,
in agreement with the theoretical prediction for the threshold singularity.

%%%%%%%%%%%%%%%%%%%%%%%%%%%%%%%%%%%%%%%%%%%%%%%%%%%%%%%%%%%%%%
%%%%%%%%%%%%%%%%%%%%%%%%%%%%%%%%%%%%%%%%%%%%%%%%%%%%%%%%%%%%%%

The universal spectral function in the MI is remarkably similar to its SF counterpart
featuring a sharp resonance peak. 
(Since MI and SF are separated by a critical line their scaling functions
$\Phi_{\rm MI}$ and $\Phi_{\rm SF}$ remain fundamentally different at energies smaller than $\omega_H$).
This observation is rather counterintuitive given 
that the superfluid order parameter is zero and raises a number of theoretical questions 
regarding the nature and properties of collective excitations in the MI phase at finite energies.
In particular, can it be linked to the established picture of renormalized free-energy functional 
for the order parameter field \cite{Berges} at distances under the correlation length?
  
If finite energy excitations probe system correlations predominantly in a finite space-time volume, 
one would expect that some resonant feature may survive even in the NL phase at sufficiently low, 
but finite temperature $T<J$ (at $g=g_{\mu}=0$, the superfluid transition temperature is zero)
In this quantum critical region, temperature determines the characteristic energy scale, thus
$S_{\rm NL}(\omega) \propto T^{3-2/\nu}\Phi_{\rm NL}(\omega /T)$, and all excitations are strongly damped.
Simulations performed at $T/J=0.5$ on the trajectory {\romannumeral 4} in the inset of Fig.~\ref{fig:phasediag}
indeed find a peak at low energies before the critical quasiplateau, see Fig.~\ref{fig:main},
but it is much less pronounced and the oscillatory component (second peak) is lost.
Unfortunately, numerical complexity does not allow us to verify the scaling law directly by
collapsing simulations at lower temperatures and bigger system sizes. Our case for universality
of $\Phi_{\rm NL}(x)$ is thus much weaker and rests solely on the theoretical consideration 
that the plateau (at the same value as in the SF and MI phases) separates universal physics from 
model specific behavior.  

In conclusion, we have constructed the universal spectral functions
$\Phi $ for the kinetic energy correlation function for all three phases
in the vicinity of the interaction driven QCP of the 2D Bose-Hubbard model.
Although the nature of excitations in these phases is fundamentally
different at low temperature, their $\Phi$ functions all feature a resonance
peak which in the SF and MI phases is followed by a broad second peak and evolve then
to a quasiplatform at higher energy in agreement with scaling predictions.
In the SF phase, the first peak is interpreted as a damped Higgs mode.
In the  MI and NL phase, the existence of a resonance is unexpected and requires 
further theoretical understanding of amplitude oscillations at mesoscopic length scales.  
Experimental verification with cold gases requires flatter traps and lower temperatures
and is accessible within current technology. It would signify a new hallmark, going
beyond the previous studies of criticality near Gaussian fixed points.

We wish to thank I. Bloch, M. Endres, D. Podolsky, and B. V. Svistunov for valuable discussions. This work was supported by the National Science Foundation Grant No. PHY-1005543, by a grant from the Army Research Office with funding from DARPA, and partially by NNSFC Grant No. 11275185, CAS, and NKBRSFC Grant No. 2011CB921300.

{\it Note added -- } During the final stage of this work, the authors of Ref.~\cite{Podolsky_new} shared with us their results for the SF phase based on MC simulations of a classical model belonging to the same universality class. We agree on the existence and the position of the Higgs resonance.


\begin{thebibliography}{99}

\bibitem{ruegg}
Ch. R{\"u}egg, B. Normand, M. Matsumoto, A. Furrer, D. F. McMorrow, K. W. Kr{\"a}mer, H. -U. G{\"u}del, S. N. Gvasaliya, H. Mutka, and M. Boehm,
Phys. Rev. Lett. {\bf 100}, 205701 (2008).

\bibitem{Bissbort}
U. Bissbort, S. G{\"o}tze, Y. Li, J. Heinze, J. S. Krauser, M.
Weinberg, C. Becker, K. Sengstock, and W. Hofstetter,
Phys. Rev. Lett. {\bf 106}, 205303 (2011).

\bibitem{chubukov}
A. V. Chubukov, S. Sachdev, and J. Ye,
Phys. Rev. B {\bf 49}, 11919 (1994).

\bibitem{sachdev}
%Sachdev, S. Universal relaxational dynamics near two-dimensional quantum critical points.
S. Sachdev,
Phys. Rev. B {\bf 59}, 14054 (1999).

\bibitem{zwerger}
%Zwerger, W. Anomalous Fluctuations in Phases with a Broken Continuous Symmetry.
W. Zwerger,
Phys. Rev. Lett. {\bf 92}, 027203 (2004).

\bibitem{book}
%Sachdev, S. Quantum Phase Transitions 2nd ed. (Cambridge University Press, Cambridge, 2011).
S. Sachdev, {\it Quantum Phase Transitions}, 2nd ed. (Cambridge University Press, Cambridge, 2011).


\bibitem{pod11}
%Podolsky, D., Auerbach, A. \& Arovas, D. P. Visibility of the amplitude (Higgs) mode in condensed matter.
D. Podolsky, A. Auerbach, and D. P. Arovas,
Phys. Rev. B {\bf 84}, 174522 (2011).

\bibitem{huber}
%Huber, S. D., Altman, E., Buchler, H. P. \& Blatter, G. Dynamical properties of ultracold bosons in an optical lattice.
S. D. Huber, E. Altman, H. P. B{\"u}chler, and G. Blatter,
Phys. Rev. B {\bf 75}, 085106 (2007).
%
\bibitem{huber2}
S. D. Huber, B. Theiler, E. Altman, and G. Blatter,
Phys. Rev. Lett. {\bf 100}, 050404 (2008).

\bibitem{Machta}
E. Burovski, J. Machta, N.V. Prokof'ev,  and  B.V. Svistunov, Phys. Rev. B {\bf 74} 132502 (2006).

\bibitem{Vicari} M. Campostrini, M. Hasenbusch, A. Pelissetto, and E. Vicari, Phys. Rev. B 74, 144506 (2006).


\bibitem{bloch}
%Endres, M., Fukuhara, T., Pekker, D.,  Cheneau, M., Schau$\beta$, P., Gross, C., Demler, E., Kuhr, S., \& Bloch, I.,
%The `Higgs' Amplitude Mode at the Two-Dimensional Superfluid-Mott Insulator
%Transition, submitted to Nature (2012).
M. Endres, T. Fukuhara, D. Pekker, M. Cheneau, P. Schau${\beta}$, C. Gross, E. Demler, S. Kuhr, and I. Bloch,
Nature {\bf 487}, 454-458 (2012).


\bibitem{lode_review}
L. Pollet, Rep. Prog. Phys. {\bf 75}, 094501 (2012).


\bibitem{pollet}
L. Pollet and N. Prokof'ev, Phys. Rev. Lett. 109, 010401 (2012).

%\bibitem{anderson}
%Anderson, P. W. Plasmons, Gauge Invariance, and Mass.
%P. W. Anderson,
%Phys. Rev. {\bf 130}, 439 (1963).

%\bibitem{higgs}
%Higgs, P. W. Broken Symmetries and the Masses of Gauge Bosons.
%P. W. Higgs,
%Phys. Rev. Lett. {\bf 13}, 508 (1964).

%\bibitem{standard}% Weinberg, S. The quantum theory of fields, Vol. 2,
%S. Weinberg,  {\it The quantum theory of fields}, Vol. 2, Cambridge University Press, Cambridge (1996).

\bibitem{fisher89} %Fisher,  M. P. A.,  Weichman,  P. B., Grinstein, G., \&  Fisher, D. S. Boson localization and the superfluid-insulator transition,
M. P. A. Fisher, P. B. Weichman, G. Grinstein, and D. S. Fisher,
Phys. Rev. B {\bf 40}, 546 (1989).


\bibitem{Podev}
 D. Podolsky and S. Sachdev, Phys. Rev. B {\bf 86}, 054508 (2012).

\bibitem{Berges} J. Berges, N. Tetradis, and C. Wetterich, Phys. Rept. {\bf 363}, 223 (2002).

\bibitem{worm}
N. V. Prokof'ev, B. V. Svistunov, and I. S. Tupitsyn,
%Worm Algorithm in the quantum Monte Carlo simulations,
Phys. Lett. A, {\bf 238}, 253 (1998);

\bibitem{worm2}
%Exact, complete and universal continuous-time worldline Monte Carlo approach to the statistics of discrete quantum systems,
N. V. Prokof'ev, B. V. Svistunov, and I. S. Tupitsyn,
Sov. Phys. - JETP {\bf 87}, 310 (1998).
%

\bibitem{worm_lode}
%% Pollet, L., Van Houcke, K., \& Rombouts, S.
%%Engineering Local optimality in Quantum Monte Carlo algorithms
L. Pollet, K. Van Houcke, and S. Rombouts,
Comp. Phys. {\bf 225}, 2249 (2007).

\bibitem{monien} N. Elstner, and H. Monien,
Phys. Rev. B {\bf 59}, 12184 (1999).

\bibitem{soyler}
%Monte Carlo Study of the Two-dimensional Bose-Hubbard Model,
B. Capogrosso-Sansone, S. G. S\"{o}yler, N. V.  Prokof'ev,  and B. V. Svistunov,
Phys. Rev. A {\bf 77}, 015602 (2008).

\bibitem{soyler2}
%%Phase Diagram of Commensurate Two-Dimensional Disordered Bose Hubbard Model,
S. G. S\"{o}yler,  M. Kiselev, N. V.  Prokof'ev, and B.V. Svistunov,
Phys. Rev Lett. {\bf 107}, 185301, (2011).

\bibitem{Podolsky_new}
S. Gazit, D. Podolsky, and A. Auerbach, Phys. Rev. Lett.
{\bf 110}, 140401 (2013).

\end{thebibliography}
\end{document}